\documentclass[aps,prb,twocolumn,superscriptaddress]{revtex4-2}

\usepackage{graphicx}
\usepackage{multirow}
\usepackage[version=4]{mhchem}
\usepackage{siunitx}
\usepackage{cleveref}

\newcommand{\fn}{\ce{Fe16N2}}
\newcommand{\afn}{\ce{$\alpha''$-Fe16N2}}

\begin{document}


\title{Van Hove singularity and phase instability: Exploring the role of electron correlation in the magnetic behavior of \afn}

\author{Peter Stoeckl}
\affiliation{School of Physics and Astronomy, University of Minnesota, Minneapolis, Minnesota 55455, USA}
\author{Przemyslaw Wojciech Swatek}
\affiliation{Electrical and Computer Engineering Department, University of Minnesota, Minneapolis, Minnesota 55455, USA}
\author{Jian-Ping Wang}
\email{jpwang@umn.edu}
\affiliation{School of Physics and Astronomy, University of Minnesota, Minneapolis, Minnesota 55455, USA}
\affiliation{Electrical and Computer Engineering Department, University of Minnesota, Minneapolis, Minnesota 55455, USA}

\date{June 18, 2026}

\begin{abstract}
The ordered iron nitride phase \afn\ is a promising candidate for environment-friendly, rare-earth-free permanent magnets due to its demonstrated giant saturation magnetization ($M_s$). However, first-principles electronic-structure calculations have struggled to consistently reproduce experimentally-observed high $M_s$, and have yielded highly variable magneto-crystalline anisotropy (MCA) values. In this work, we employ Density Functional Theory under the GGA+$U$ framework to study the effect of the Hubbard parameters $U$ and $J$ on the magnetic properties of \fn. We demonstrate that the electronic structure exhibits high sensitivity to these parameters, specifically uncovering a van Hove singularity near the Fermi level ($E_F$), inherently tied to the material's structural and thermal phase instability. By linking this topological anomaly to the calculated magnetic properties, we demonstrate that the selection of $U$ not only tunes $M_s$ and MCA energy towards experimental values but also reveals an underlying electronic mechanism potentially responsible for the phase's metastability. This provides a framework for understanding the correlation-driven magnetic behavior of \fn\ and offers a pathway for  optimizing its stability and performance in practical applications.
\end{abstract}


\maketitle

\section{Introduction}

The ordered iron nitride phase \afn\ stands out as one of the most promising candidates for next-generation environment-friendly, rare-earth-free permanent magnets.\cite{wang2020} First discovered by Jack in 1951,\cite{jack1951} the material garnered immense excitement from initial reports of a ``giant'' saturation magnetization ($M_s$) in 1972 and 1991.\cite{kim1972,sugita1991} These values exceeded the maximum theoretical limit for conventional itinerant ferromagnets like Fe--Co alloys, suggesting its use for high-density magnetic recording and power applications. However, this early promise was hampered by subsequent experimental studies through the mid-1990s that yielded inconclusive results.\cite{coey1994b} The renewed, intense global interest is primarily driven by recent, reproducible observations of high $M_s$ samples,\cite{ji2011a,wang2012,ji2013} particularly the seminal work in advanced thin-film synthesis and characterization which confirmed a giant saturation magnetic flux density of up to \qty{2.9}{T}, a value unmatched by other rare-earth-free materials.

Attempts to theoretically reproduce this high $M_s$ through density-functional theory (DFT) have largely remained inconclusive. Conventional methods including the common generalized gradient approximation (GGA) often fail to replicate the experimentally observed higher magnetization values.\cite{coey1994b} To bridge this theoretical gap and gain insight into potential mechanisms, several more advanced paradigms have been employed. In particular, the addition of the Hubbard correction term $U$ (and Hund's term $J$) to account for strong electron-electron correlations in the Fe(3d) shell has been increasingly considered.\cite{lai1994,ji2010,sims2012,devi2022} These correlation-corrected calculations often suggest that the giant moment arises from an unusual effect leading to a partial localization of the 3d electron states in the Fe--N clusters.\cite{ji2010}

In permanent (hard) magnet applications, high coercivity is essential in addition to high $M_s$. This resistance to demagnetization is microscopically associated with the magneto-crystalline anisotropy (MCA),\cite{wang2020} which is the energy difference between the easy and hard axes of magnetization. The \afn\ phase exhibits uniaxial anisotropy, but experimental data reports a wide range of anisotropy constants, $K_u \sim \qtyrange{4.4e6}{1.9e7}{erg/cc}$,\cite{takahashi1999,kita2007,ji2011b,ogawa2013,li2019,liu2020} which is generally lower than the target range for strong permanent magnets ($\sim \qtyrange{5e7}{2e8}{erg/cc}$).\cite{weller2000} Theoretical calculations show a similar range\cite{li2019,ke2013,zhou2014,zhao2016,han2019,islam2020,sun2020,
ochirkhuyag2021a,kota2023,sakuma2023} (compare also \Cref{t:sm_prev_mca} in the Supplemental Material.\cite{sm_note}

\begin{table*}
 \caption{\label{t:prev_calc} Selected previous calculations of MCA energy in \fn\ using DFT+$U$, including choices of parameters and lattice constants used for context. Where one value is given for the Hubbard constants, it corresponds to $U_{eff}$  (or $U-J$); where two, to $U, J$, respectively.
}
 \begin{ruledtabular}
  \begin{tabular}{c c c ccc c cc c}
  \multicolumn{2}{c}{Source} && \multicolumn{3}{c}{Hubbard $U_{eff}$ [or $U/J$] (eV)} && \multicolumn{2}{c}{Lattice Const. (\AA)} & $K_u$ \\
  \cline{1-2}\cline{4-6}\cline{8-9}
  Paper & Notes && Fe$4e$ & Fe$8h$ & Fe$4d$ && $a$ & $c$ & ($10^5$ erg/cc)\\
  \hline
  \multirow{2}{*}{Zhou 2014\cite{zhou2014}} & relaxed lattice && 3.12\,/\,0.59 & 3.52\,/\,0.61 & 3.99\,/\,0.64 && 5.68 & 6.23 & 146 \\
 & in-plane strain && 3.12\,/\,0.59 & 3.52\,/\,0.61 & 3.99\,/\,0.64 && 5.91 & 6.23 & -8.8 \\
  \hline
  \multirow{3}{*}{Li 2019\cite{li2019}} & exp. lattice && 1.0 & 1.0 & 1.0 && 5.71 & 6.28 & 50 \\
 & exp. lattice && 4.0 & 4.0 & 4.0 && 5.71 & 6.28 & 160 \\
 & strain along $c$ && 4.0 & 4.0 & 4.0 && 5.71 & 6.61 & 210 \\
  \hline
Islam\&Borah & \multirow{2}{*}{relaxed lattice} && \multirow{2}{*}{1.088} & \multirow{2}{*}{1.360} & \multirow{2}{*}{3.944} && \multirow{2}{*}{5.696} & \multirow{2}{*}{6.185} & \multirow{2}{*}{81} \\
 2020\cite{islam2020} & & && & & && & \\
  \end{tabular}
 \end{ruledtabular}
\end{table*}

Previous calculations of MCA energy in \fn\ beyond the GGA paradigm have predominantly employed GGA+$U$. The selection of appropriate values for the Hubbard $U$ and $J$ parameters is a known point of complexity; these values are generally not transferable between different systems or DFT implementations.\cite{cococcioni2005} As a result, the broad range of choices for $U$ and $J$ has led to a variety of estimates for MCA energy in \fn, including in our own previous work (see also \Cref{t:prev_calc}).\cite{li2019,zhou2014,islam2020,stoeckl2021,stoeckl2023}

In this work, we investigate more comprehensively the effect of Hubbard $U$ and $J$ parameters on the magnetic properties of \afn, particularly focusing on the magnetization and MCA energy. Furthermore, we will link the selection of these correlation parameters to i) high magnetization, ii) MCA energy, and iii) phase instability by observing anomalous electronic features observed near the Fermi level ($E_F$), a van Hove singularity, demonstrating how the choice of $U$ can directly influence the theoretical prediction of material stability and the resulting experimental magnetic properties.


\section{Computational Methods}

Electronic structure calculations were performed using the plane-wave “pwscf” code from the Quantum \textsc{Espresso} package.\cite{giannozzi2009,giannozzi2017} The GGA exchange-correlation functional, as formulated by Perdew, Burke, and Ernzerhof,\cite{perdew1996} and Projector-Augmented Wave (PAW) pseudopotentials (PP) taken from the QE PSlibrary,\cite{dalcorso2014} were used throughout. Scalar-relativistic or fully-relativistic PP variants were employed as necessary. In GGA+$U$ calculations, due to technical limitations, two variants were used: a simplified formulation with effective Hubbard correction $U_{eff} \sim U-J$,\cite{cococcioni2005} and a more explicit formulation using both $U$ and $J$ terms due to Liechtenstein \textit{et al.}\cite{liechtenstein1995}

In structural relaxations, the primitive (9-atom) unit cell's volume, shape, and atomic positions were allowed to relax within a scalar-relativistic approach, neglecting spin-orbit coupling (SOC), until atomic forces were below \qty{0.05}{eV/\textup{\AA}} and stress tensor elements below \qty{0.5}{kbar} (\qty{0.05}{GPa}). For GGA+$U$, the simplified $U_{eff}$ formulation was used. Calculations were performed with kinetic energy cutoff of \qty{150}{Ry} ($\sim \qty{2000}{eV}$); the Brillouin zone (BZ) was sampled via the Monkhorst-Pack scheme with an $8\times 8\times 8$ grid, with Marzari-Vanderbilt cold smearing of width \qty{0.14}{eV}. 

MCA energy calculations were performed using the respective relaxed structures for each choice of $U$ and $J$ (or pure GGA), as MCA is particularly sensitive to lattice strain.\cite{ke2013,zhou2014} For these calculations, $k$-point mesh density increased to $24 \times 24 \times 24$, with smearing narrowed to \qty{0.07}{eV}. SOC terms (and fully-relativistic PP) were included; for GGA+$U$, the explicit $U+J$ method was employed. The MCA energy was calculated as the difference in total energy between states with magnetization along the easy (001) and hard (100) directions, i.e. $
\Delta E_{MCA} = E_(100) - E_(001)$, scaled to the conventional (18-atom) unit cell; magnetic moments (similarly scaled) were taken from the calculation with (001) spin.

Projected density of states (pDOS) was calculated in the ground state ($M$ along the easy axis) without SOC, with the finer $k$-mesh, explicit $U+J$ method, and the tetrahedron method due to Bl\"{o}chl.\cite{blochl1994} Local moments and net charges were determined by L\"{o}wdin population analysis. Band structures used an intermediate $k$-mesh of $16 \times 16 \times 16$, followed by plotting $k$-dependent pDOS to determine orbital character, using a narrow \qty{0.03}{eV} Gaussian broadening instead of cold smearing. Fermi surfaces were obtained with the fine $k$-mesh shifted to center on the origin, with 3D plots and 2D sections of the BZ generated using the FermiSurfer code.\cite{kawamura2019}


\section{Results and Discussion}

\subsection{Initial exploration}

\begin{figure}
 \includegraphics{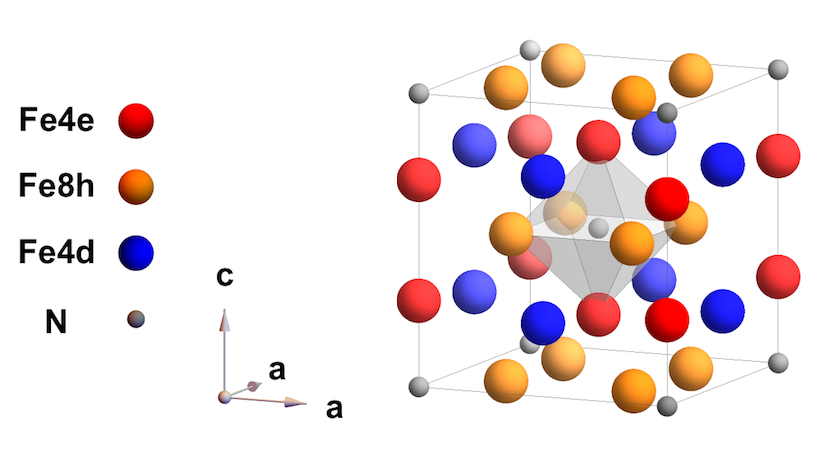}%
 \caption{\label{f:structure} Crystal Structure of \afn, showing the bct unit cell, its axes, and specific Fe Wyckoff positions. (Figure produced using \textsc{Mathematica}, Version 11.3.0, by Wolfram Research, Inc.\cite{Mathematica2018})}
\end{figure}

The experimental crystal structure of \afn\ is body-centered tetragonal, with lattice constants $a = \qty{5.72}{\textup{\AA}}$ and $c = \qty{6.29}{
\textup{\AA}}$, and Fe atoms in three different Wyckoff positions: $4e$ and $8h$ (adjacent to the interstitial N atoms), and $4d$.\cite{jack1951} See \Cref{f:structure}.

Our first survey of GGA+$U$ parameters was previously reported in 2023;\cite{stoeckl2023} for context, we provide an abbreviated summary of these results here, with additional detail in the Supplemental Material including \Cref{f:sm_uj_data,f:sm_j_data}.\cite{sm_note} Following the model proposed by Ji \textit{et al.} in 2010,\cite{ji2010} Hubbard $U$ was fixed at a low value of \qty{1.0}{eV} on the Fe$4d$ site while simultaneously varying $U$ on the Fe$4e$ and Fe$8h$ sites. In calculations with SOC, the Hund's coupling parameter was taken to be $J=U/10$ on all three Fe sites. For the preceding structural relaxation calculations, the effective parameter $U_{eff}$ was set equal to $U-J$ from the SOC calculations.

The following trends were observed: Increasing $U-J$ on Fe$4e$/$8h$ mainly led to expansion in the $ab$ plane, increasing unit cell volume while reducing $c/a$. While magnetic moment increased with greater $U-J$, saturation magnetization $M_s$ exhibited a peak due to lattice expansion. MCA energy, however, behaved less linearly, including a change in sign (from out-of-plane to in-plane easy axis) before returning to positive values as $U-J$ increases.

To minimize the potential discrepancies between the structural relaxation calculations (using $U_{eff}$) and the magnetic property calculations (using $U$ and $J$), a second systematic survey was conducted with $J=0$ throughout, so that the $U$ value used in the SOC calculations matched $U_{eff}$ used for the structural relaxation (as $U_{eff} = U - J$ now simplifies to $U_{eff} = U$). In this series, the trends were similar; however, the region of decreased or negative MCA energy was greatly reduced, implying that this nonlinear behavior cannot be entirely explained by the ``effective strain'' resulting from using $U_{eff}$ to relax crystal structures rather than the explicit $U,J$ used in magnetic property calculations.

\subsection{Comprehensive Hubbard $U$ landscape}

\begin{figure*}
 \includegraphics{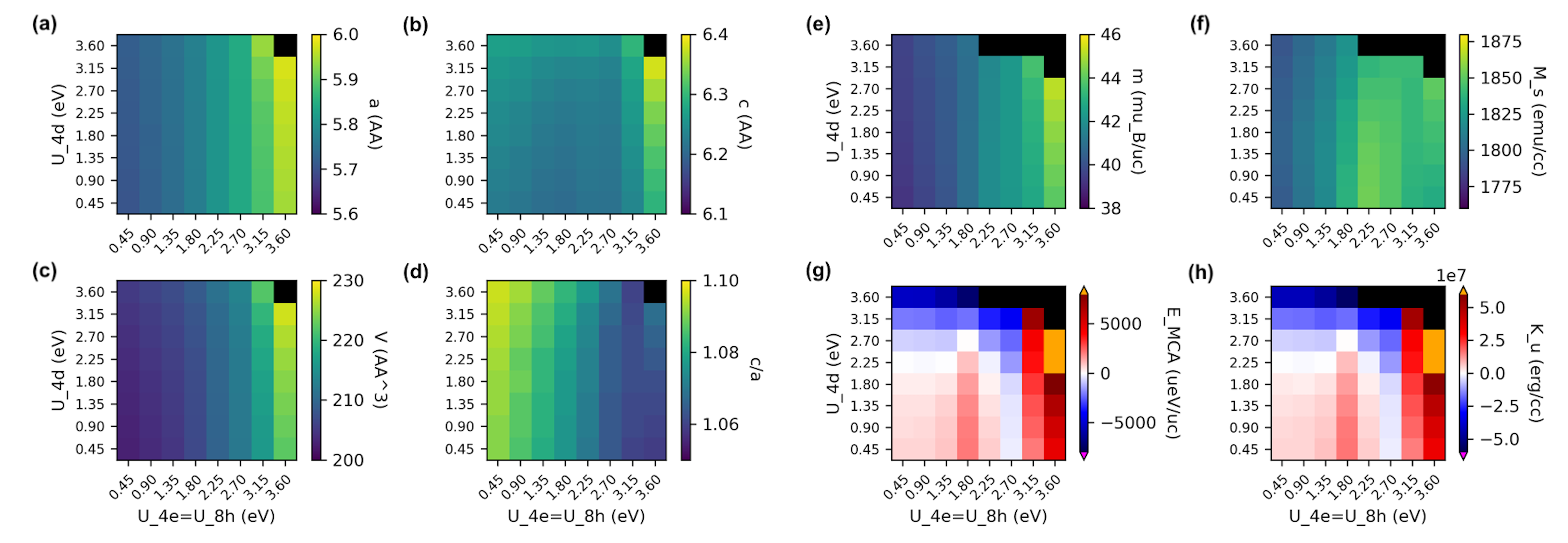}%
 \caption{\label{f:set_eq} Lattice and magnetic parameters of \fn\ under GGA+$U$, with $U_{4e}=U_{8h}$ varying independently of $U_{4d}$. (a--d) Lattice parameters. (e--h) Magnetic parameters. Black squares indicate missing data.}
\end{figure*}

\begin{figure*}
 \includegraphics{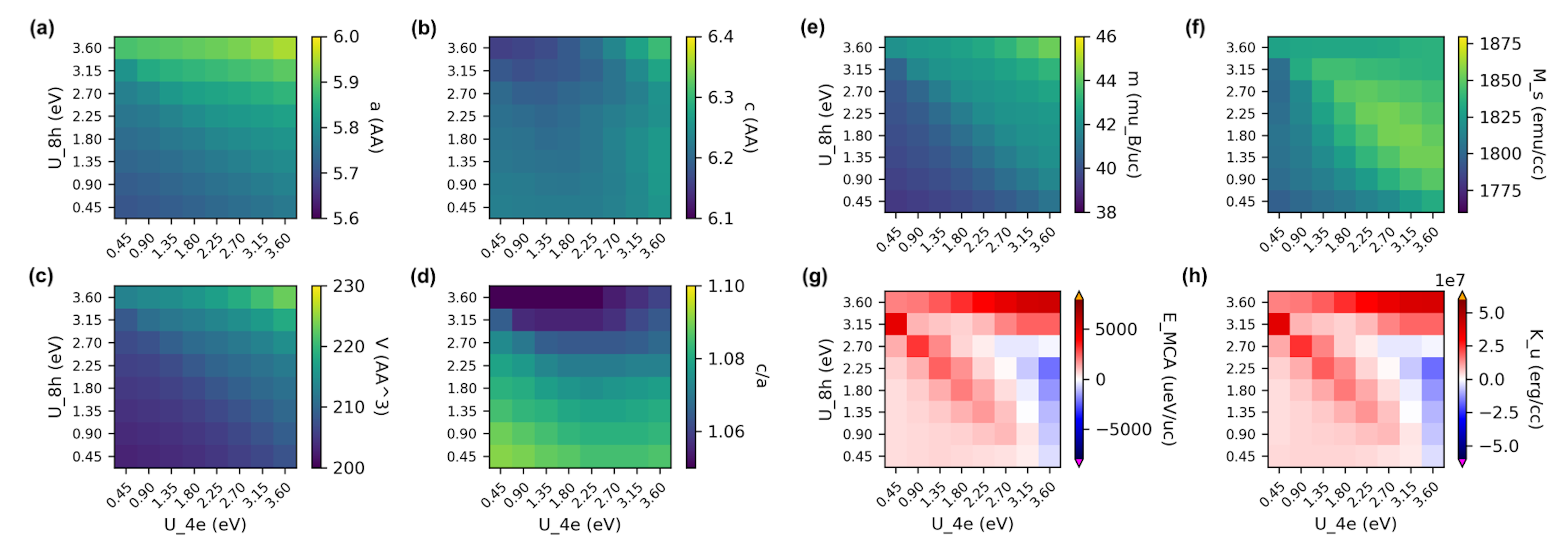}%
 \caption{\label{f:set_uneq} Lattice and magnetic parameters of \fn\ under GGA+$U$, with $U_{4e}$ and $U_{8h}$ varying independently, while $U_{4d}=\qty{0.90}{eV}$ fixed. (a--d) Lattice parameters. (e--h) Magnetic parameters. Black squares indicate missing data.}
\end{figure*}

To obtain a more comprehensive picture of the parameter landscape, including coverage of parameter ranges used by previous work on GGA+$U$ on \fn, we obtain data for a regular grid of Hubbard $U$ values (taking $J=0$ throughout for reasons discussed above, and maintaining $U$ value spacing of \qty{0.45}{eV} for continuity with the previous series). First, we consider the case where $U_{4e} = U_{8h}$, so both atoms adjacent to N are treated identically, while $U$ on Fe$4d$ varies independently (see \Cref{f:set_eq}). Some data points are absent, particularly at higher $U$ values, due to non-convergence of the self-consistent field (SCF) cycles. This limitation arises when the resulting electronic configuration leads to highly localized states that exceed the maximum tolerance of our computational setup, thus defining the upper bound of the parameter space that can be reliably explored.

For relaxed lattice parameters, the previously observed dependence of $a$ on $U_{4e}$ and $U_{8h}$ persists in the full dataset, while $c$ increases with all three $U$ parameters. As before, the $a$ value exceeds the experimental lattice parameter at moderate $U$, whereas $c$ remains below experiment until $U$ is well above 3 eV. Magnetic moments similarly increase with either $U$ parameter, though more strongly with $U_{4e}=U_{8h}$; again, while moments continue to increase with $U$, $M_s$ peaks at lower $U$ values due to lattice expansion. This peak value $\sim\qty{1850}{emu/cc}$ remains well below several recent experimental observations of $M_s > \qty{2000}{emu/cc}$.\cite{wang2012,jiang2016c,feng2017,zhang2017,chen2019,hang2020} Finally, anisotropy remains highly variable: at low $U_{4d}$, we observe the same behavior as before, with MCA energy first increasing, then changing sign, then increasing again up to $U=\qty{3.60}{eV}$. As $U_{4d}$ increases, however, anisotropy tends to decrease, with MCA energy strongly negative (easy axis in-plane) at high $U_{4d}$ and low to moderate $U_{4e}=U_{8h}$. When all three $U$ parameters are high, meanwhile, MCA energy is once again strongly positive and increases further with $U$.

Next, we consider the effect of varying $U_{4e}$ and $U_{8h}$ independently, as illustrated for fixed $U_{4d} = \qty{0.90}{eV}$ by \Cref{f:set_uneq}. The lattice parameters behave as expected given the configuration of Fe around N: $a$ depends most strongly on $U$ on the in-plane Fe$8h$, while $c$ depends most strongly on the vertically-oriented Fe$4e$. Magnetic moments, meanwhile, increase similarly with either $U_{4e}$ or $U_{8h}$. As for MCA energy, this is again strongly variable; here, anisotropy peaks along the antidiagonal (from low $U_{4e}$/high $U_{8h}$ to the opposite), as well as when both $U$ parameters are high, while for high $U_{4e}$ and low to moderate $U_{8h}$ anisotropy is again negative. Similar behavior can be observed at higher fixed $U_{4d} = 1.80, 2.70$ (see SM, \Cref{f:sm_set_uneq2,f:sm_set_uneq3}\cite{sm_note}).

\subsection{Detailed electronic structure}

\begin{table*}
 \caption{\label{t:cases} Selected cases of interest: Hubbard U parameters and resultant magnetic properties. \textbf{A0} ($U=0$ throughout) is equivalent to pure GGA.
}
 \begin{ruledtabular}
  \begin{tabular}{c ccc r r r r}
  \multirow{2}{*}{Label} & \multicolumn{3}{c}{Hubbard $U$ (eV)} & Net moment & Equiv. $M_s$ & $\Delta E_{MCA}$ & Equiv. $K_u$ \\
  \cline{2-4}
  & Fe$4e$ & Fe$8h$ & Fe$4d$ & ($\mu_B$/cell) & (emu/cc) & ($\mu$eV/cell) & ($10^5$ erg/cc) \\
  \hline
  \textbf{A0} & 0 & 0 & 0 & 38.30 & 1766 & +909 & +72.3\\
  \textbf{A1} & 1.80 & 1.80 & 0.45 & 40.92 & 1837 & $+2113$ & $+163.9$\\
  \textbf{A2} & 0.45 & 0.45 & 3.15 & 39.58 & 1794 & $-2117$ & $-165.7$\\
  \textbf{A3} & 1.35 & 1.35 & 2.25 & 40.26 & 1815 & $-25.6$ & $-2.0$\\
  \textbf{A4} & 3.15 & 3.15 & 2.25 & 43.18 & 1840 & $+3795$ & $+279.3$\\
  \textbf{A5} & 3.60 & 3.60 & 0.90 & 44.20 & 1837 & $+5475$ & $+393.2$\\
  \hline
  \textbf{B1} & 3.60 & 2.25 & 0.90 & 42.30 & 1847 & $-2244$ & $-169.3$\\
  \textbf{B2} & 2.25 & 2.70 & 0.90 & 42.14 & 1850 & $+282.1$ & $+21.4$\\
  \textbf{B3} & 0.45 & 3.15 & 0.90 & 40.54 & 1804 & $+4733$ & $+363.7$\\
  \end{tabular}
 \end{ruledtabular}
\end{table*}

For greater insight into the electronic structure and the effect of Hubbard $U$, we consider the local charges and magnetic moments as well as site- and orbital-projected DOS. For ease of reference, we apply short labels to a representative selection of cases covering a wide range of values for both $U$ and resulting magnetic properties, as listed in \Cref{t:cases}; cases starting with `A' have $U_{4e}=U_{8h}$, while those starting with `B' correspond to varying $U_{4e}$ and $U_{8h}$ with $U_{4d}=\qty{0.90}{eV}$. (Case \textbf{A5} bridges both categories.)

\begin{table*}
 \caption{\label{t:moments} Local atomic charges/moments for selected cases. Note that positive charge transfer corresponds to increased negative charge.
}
 \begin{ruledtabular}
  \begin{tabular}{c cccc c cccc r}
  \multirow{2}{*}{Label} & \multicolumn{4}{c}{Charge Transfer (e)} && \multicolumn{4}{c}{Magnetic Moment ($\mu_B$)} &  $\Delta E_{MCA}$  \\
  \cline{2-5}\cline{7-10}
  & Fe$4e$ & Fe$8h$ & Fe$4d$ & N && Fe$4e$ & Fe$8h$ & Fe$4d$ & N & ($\mu$eV/cell)\\
  \hline
  \textbf{A0} & $-0.0183$ & $-0.0483$ & $-0.0877$ & $+0.2269$ && $+2.1510$ & $+2.3394$ & $+2.8364$ & $-0.0821$ & $+909$\\
  \textbf{A1} & $-0.0183$ & $-0.0625$ & $-0.0740$ & $+0.2562$ && $+2.4352$ & $+2.5807$ & $+2.7948$ & $-0.0988$ & $+2113$\\
  \textbf{A2} & $-0.0113$ & $-0.0348$ & $-0.1250$ & $+0.2351$ && $+2.2070$ & $+2.3798$ & $+3.0340$ & $-0.0960$ & $-2117$\\
  \textbf{A3} & $-0.0142$ & $-0.0442$ & $-0.1069$ & $+0.2409$ && $+2.3176$ & $+2.4717$ & $+2.9453$ & $-0.1116$ & $-25.6$\\
  \textbf{A4} & $-0.0168$ & $-0.0718$ & $-0.0815$ & $+0.3120$ && $+2.6492$ & $+2.7288$ & $+2.8986$ & $-0.0686$ & $+3795$\\
  \textbf{A5} & $-0.0316$ & $-0.0853$ & $-0.0516$ & $+0.2527$ && $+2.7633$ & $+2.8232$ & $+2.8242$ & $-0.0986$ & $+5475$\\
  \hline
  \textbf{B}1 & $-0.0399$ & $-0.0599$ & $-0.0724$ & $+0.2884$ && $+2.6921$ & $+2.6082$ & $+2.7992$ & $-0.0842$ & $-2244$\\
  \textbf{B2} & $-0.0090$ & $-0.0761$ & $-0.0677$ & $+0.2817$ && $+2.4947$ & $+2.6809$ & $+2.7901$ & $-0.0818$ & $+282$\\
  \textbf{B3} & $+0.0270$ & $-0.0878$ & $-0.0669$ & $+0.2553$ && $+2.0359$ & $+2.7357$ & $+2.7984$ & $-0.1015$ & $+4733$\\
  \end{tabular}
 \end{ruledtabular}
\end{table*}

In general, L\"{o}wdin charge transfer analysis finds that on net, partial electronic charge transfers from all three Fe sites to N (with the exception of case \textbf{B3}, which with $U_{4e}<U_{8h}$ sees net charge transfer to Fe$4e$ as well as N); compare \Cref{t:moments}. Magnetic moments on Fe range from around \qtyrange{2.3}{2.9}{\mu_B}, with $m_{4e}<m_{8h}<m_{4d}$ consistently throughout; net moment on N is weakly negative, around \qty{-0.1}{\mu_B}. As $U$ on a Fe site increases, more partial charge tends to transfer away from that site (usually to the benefit of N), and net moment increases.

\begin{figure*}
 \includegraphics{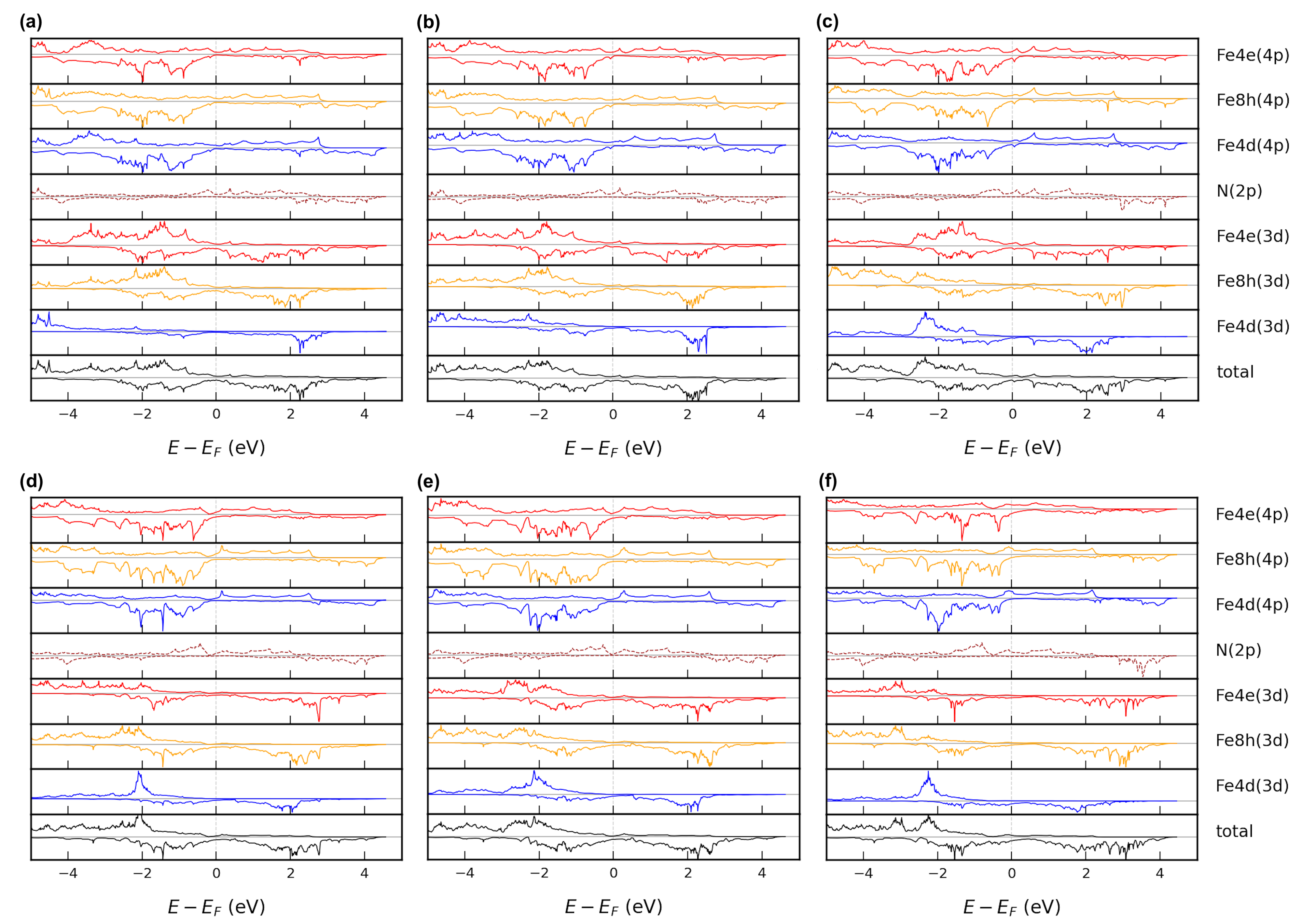}%
 \caption{\label{f:pdos_set} Site/orbital-resolved pDOS for selected illustrative cases: (a) \textbf{A2}; (b) \textbf{A3}; (c) \textbf{B3}; (d) \textbf{B1}; (e) \textbf{B2}; (f) \textbf{A5}. $\Delta E_{MCA}$ increases from left (negative) to right (positive), while $m_{avg}$ increases from top to bottom. Upper curve is spin-up, lower curve (negated) spin-down density. Orbital-resolved DOS are independently normalized to better show detail.
}
\end{figure*}

Selected pDOS plots are given in \Cref{f:pdos_set}, including spin-polarized projection onto relevant valence orbital characters. In typical fashion, 3d-shell spin-splitting tends to increase with local $U$, with spin-up density shifting down and spin-down shifting up in energy, consistent with the increased local moments. Notably, the projected Fe(4p) densities are nonzero, notwithstanding the canonical Fe electronic configuration of $4\mathrm{s}^{2}3\mathrm{d}^{6}4\mathrm{p}^{0}$. The Fe(4p) density consistently exhibits a peak in the spin-up channel located just above the Fermi energy. In general, for configurations with higher MCA energy and/or higher local moments this peak tends to sharpen and move lower in energy, i.e. closer to $E_F$, with the high-moment/high-anisotropy cases \textbf{A4} and \textbf{A5} showing the peak just above or just below $E_F$, respectively. In such instances, this sharp feature takes on the characteristics of a van Hove singularity, indicating a critical topological change in the electronic structure near the Fermi level; such features have also been linked to strong electron correlations as well as phase instability.\cite{fawcett1994,szytula2014,fujita2023,oleary2024,blawat2025}

\subsection{Band structure, Fermi surface, and strain dependence}

To further elucidate the electronic structure properties, particularly the effect of the Fe(4p) spin-up contributions on the MCA, we focus specifically on the highly anisotropic case \textbf{A4} ($U_{4e}=U_{8h}=\qty{3.15}{eV}, U_{4d}=\qty{2.25}{eV}$), with relaxed lattice constants $a = \qty{5.90}{\textup{\AA}}$ and $c = \qty{6.26}{\textup{\AA}}$, and magnetic properties $M = \qty{43.18}{\mu_B/u.c.}$ and $\Delta E_{MCA} = \qty{+3795}{\mu{}eV/u.c.}$

The Fe(4p) peak position’s relative insensitivity to $U$ (varying by $<\qty{1}{eV}$ while $U$ varies by $>\qty{3}{eV}$) suggests an intrinsic, structural origin related to the local crystal field splitting. With \afn\ exhibiting $D_{4h}$ symmetry (I4/mmm crystal structure), we therefore probe this effect by considering the electronic structure under the application of $\pm 4\%$ strain in the $ab$-plane (\textit{i.e.}, varying the lattice parameter $a$ while keeping $c$ unchanged). This strain strongly affects the magnetic properties: under compressive strain ($-4\%$),  $\Delta E_{MCA}$ changes sign to $-\qty{258}{\mu{}eV/u.c.}$, and the net moment decreases to \qty{40.16}{\mu_B/u.c.} Under expansive strain ($+4\%$), $\Delta E_{MCA}$ increases sharply to \qty{+14176}{\mu{}eV/u.c.}, with moment increasing to \qty{45.32}{\mu_B/u.c.} The change in moment agrees with previous observations by Zhou \textit{et al.} (2014);\cite{zhou2014} note however that this work observed sign change in MCA energy decreasing under tensile strain, the opposite direction of that found here.

\begin{figure*}
 \includegraphics{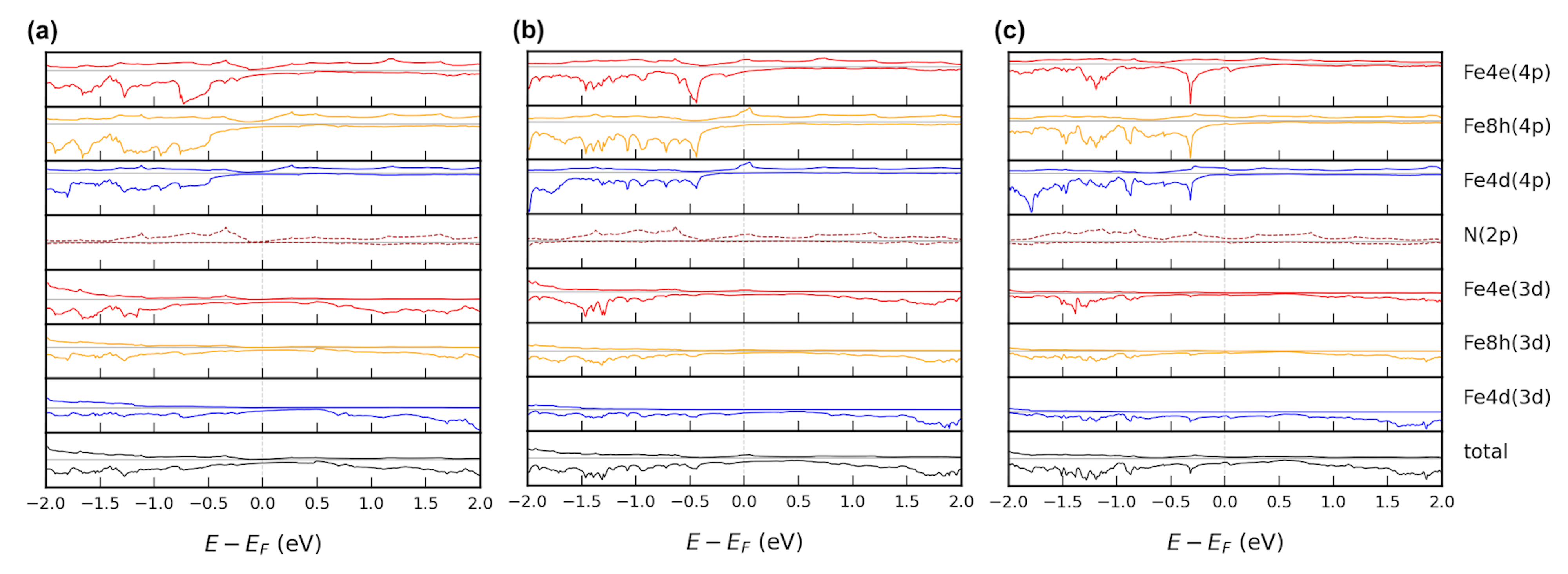}%
 \caption{\label{f:pdos_strain} Orbital-resolved pDOS near the Fermi energy, case \textbf{A4}. (a) $-4\%$ $ab$-strain; (b) 0 strain; (c) $+4\%$ $ab$-strain.
}
\end{figure*}

Projected Density of States (pDOS) of case \textbf{A4} under varying strain are shown in \Cref{f:pdos_strain}. Under zero strain, pronounced features from Fe$8h$ and Fe$4d$ are observed immediately above the Fermi level. (A similar but weaker feature is observed for N(2p) DOS, while Fe$4e$ does not appear to contribute.) Further orbital-specific projected DOS (see SM, \Cref{f:sm_crystal_field}\cite{sm_note}) show that the Fe(4p) spin-up peak can be attributed specifically to $\mathrm{p_x}/\mathrm{p_y}$ on Fe$8h$ and $\mathrm{p_z}$ on Fe$4d$. The crucial observation is that the sharp Fe(4p) spin-up peaks shift from being located slightly above $E_F$ at $-4\%$ strain to below $E_F$ at $+4\%$ strain; in both cases, the peak becomes significantly more diffuse under strain.

\begin{figure*}
 \includegraphics{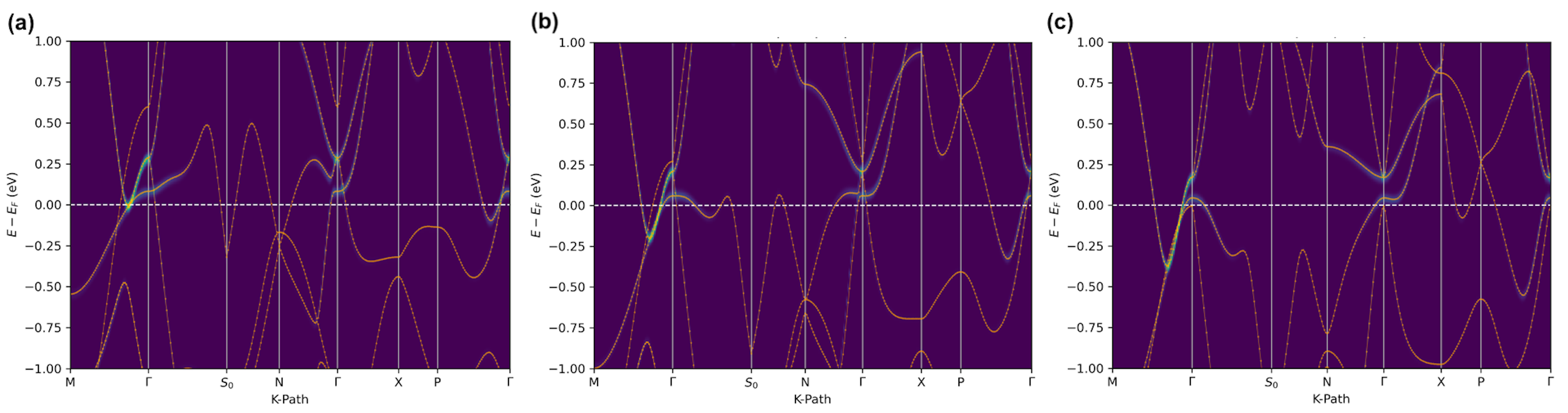}%
 \caption{\label{f:bs_strain} \emph{Spin-up} band structure (orange) near the Fermi energy for case \textbf{A4}, superimposed on $k$-resolved pDOS projected on Fe(4p) charge density (color). (a) $-4\%$ $ab$-strain; (b) 0 strain; (c) $+4\%$ $ab$-strain.
}
\end{figure*}

The corresponding spin-up band structures, plotted via $k$-resolved pDOS in \Cref{f:bs_strain}, clearly demonstrate the dominating contribution of $4p$ orbitals to the van Hove anomaly. Note the comparatively quasi-flat bands along the $\Gamma-S_0$ and $\Gamma-X$ lines in the basal plane of the BZ, capturing symmetry-equivalent directions due to the fourfold rotational symmetry ($C_4$), with dominating contribution from Fe$4d$ and Fe$8h$ sites. These bands tend to flatten under zero strain (full $\mathrm{p_x}/\mathrm{p_y}$ degeneration), creating van Hove anomalies located at the Fermi level.

To analyze the microscopic origin of these flat bands and their connection to macroscopic magnetic properties, we look to the underlying lattice symmetry. Due to the E double-degenerate representation in the $D_4h$ point group symmetry, the $\Gamma-S_0$ path reveals degenerate bands that are crucial for further analyzing Fermi surface nesting. Under biaxial strain, the $E$ set retains its in-plane rotational degeneracy due to the preserved fourfold symmetry. The small feature observed at the Fermi level for zero strain moves toward either lower or higher energy under the applied pressure, lifting the degeneracies along $\Gamma-S_0$ and leading to band splitting near $E_F$. This provides direct evidence that the MCA and magnetization in \fn\ are highly sensitive to a combination of the crystallographic parameters, crystal field splitting, and thus the filling of the Fe(4p) bands near $E_F$.

Interestingly, while the local Fe(4p) bands split and shift under strain, the weaker peak in N(2p) pDOS shifts similarly but remains fairly unchanged in shape; projected band structures indicates that this N(2p) feature corresponds to different bands than the Fe(4p) peak discussed above (Supplemental Material, \Cref{f:sm_pdos_n2p}\cite{sm_note}). This points to a rather weak sensitivity to direct covalent orbital hybridization. It appears, therefore, that the dominant factor influencing this highly sensitive localized electronic structure is the geometry of the crystal structure and the strength of the local crystal field rather than changes in the intraelectronic interactions associated with the Coulomb potential $U$ itself.

Nevertheless, a small anomaly localized at the Fermi level at zero strain can lead to significant electronic instability when combined with the interaction of these highly polarized Fe(4p) orbitals. Based on this, we can assume that the highly unstable character of the pure relaxed \fn\ phase (in this high magnetic and high-MCA energy case) is associated with this van Hove anomaly, driven mostly by the filling and band hybridization of Fe(4p) orbitals, with moderate or minor contribution from N(2p) orbitals. These results align well with the experimental reports indicating that doping via other p-, d-elements, 1D/2D material growth, or introducing defects to the lattice is required for \fn\ stabilization at room temperature,\cite{wang1999b,takahashi2000,wang2002a,jiang2016c,
zhang2013,jiang2016a,ma2020} which in practice imply modification of electron structure in the vicinity of the Fermi level.

Finally, we investigate the Fermi surface topology to visually map the electronic consequences of the shifting flat bands. The spin-up surfaces under $-4\%$, $0\%$, and $+4\%$ strain are shown, divided by band index, in SM, \Cref{f:sm_fs_strain}.\cite{sm_note} The sheet belonging to the primary contributing band undergoes significant morphological changes under applied strain, with smaller, higher-energy bands becoming increasingly relevant at $+4\%$ strain.
 
\begin{figure*}
 \includegraphics{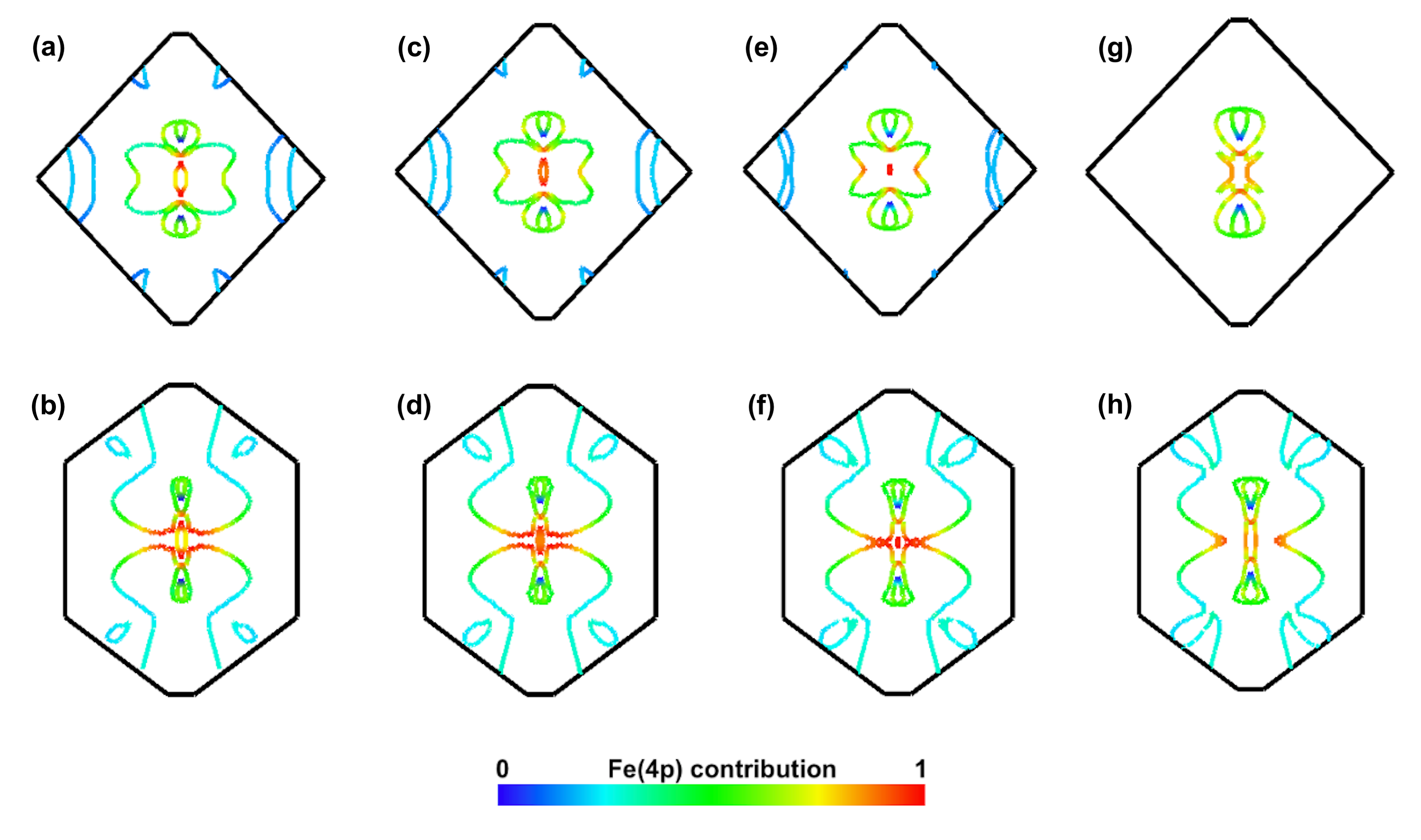}%
 \caption{\label{f:fs_vhs} Constant energy cuts through the BZ (\textit{spin up} only) for case \textbf{A4}, at $E - E_F$ (a--b) \qty{0}{eV} (Fermi surface proper); (c--d) \qty{0.025}{eV}; (e--f) \qty{0.050}{eV}; (g--h) \qty{0.075}{eV}. Cuts pass through (100), $xz$-plane (a,c,e,g); (110) plane (b,d,f,h). Color indicates Fe(4p) projection.
}
\end{figure*} 
 
For further insight, the evolution of the Fermi surface topology with chemical potential (at zero strain) is shown through the 2D sections in \Cref{f:fs_vhs}; the critical Fe(4p) contribution is highlighted. As the chemical potential is tuned through the van Hove singularity, a clear Lifshitz transition is observed.\cite{liu2010,wu2015,swatek2018} The Fermi surface pockets undergo a critical reconnection, characterized by a transition from closed orbits to open, saddle-point-dominated contours between \Cref{f:fs_vhs}(c--d) and \Cref{f:fs_vhs}(e--f). Near this anomaly, the flattening of the Fermi arcs leads to a significant enhancement of Fermi surface nesting, where large parallel segments are connected by a nesting vector $\mathbf{q}$.

This confluence of a divergent DOS and strong nesting suggests a platform for the emergence of many-body instabilities. It is widely known that materials hosting a sharp van Hove singularity near the Fermi level show a variety of profound spin-fluctuation features or many-body instabilities. To minimize the electronic free energy, these fluctuations can be removed by symmetry-breaking mechanisms, such as a structural phase transition, the onset of superconductivity, or additional non-commensurate magnetic ordering driven by local band dispersion modifications. In our system, the lifting of these degeneracies alters the band topology near $E_F$, directly enhancing Fermi surface nesting and potentially enhancing spin polarization.

Given the ferromagnetic nature of \fn, these effects are likely to manifest as instability of the magnetic structure, potentially triggering spin density waves or complex non-collinear ordering.\cite{fawcett1994,szytula2014,fujita2023,oleary2024,blawat2025} Furthermore, the electronic coupling to the lattice near these saddle points may drive structural instabilities or lattice reconstructions---such as a Peierls-like distortion---as the system seeks to lower its total energy by lifting the degeneracy at the singularity. This structural susceptibility may explain the reduced thermal stability of purely ordered \afn\ at room temperature,\cite{wang2020,stoeckl2025} while offering further opportunities for material tuning. Most critically for the magnetocrystalline anisotropy, if the hybridization between Fe-d and N-p states is sensitive to the van Hove anomaly, the system may favor specific orbital ordering to break the symmetry.

For comparison, similar investigations for two other high-anisotropy cases (\textbf{A5} and \textbf{B3}) are provided in the Supplemental Material (\Cref{f:sm_case_a5,f:sm_case_b3}).\cite{sm_note} Case \textbf{A5} has similar strain-dependence of magnetic moment and anisotropy, both increasing with lattice $a$. The Fe(4p) peak shows similar shifts, though here it is most well-defined at $-4\%$ strain; the correlation of this peak to specific flattened bands is less strong here than in case \textbf{A4}.

Case \textbf{B3} however, with lower magnetic moment (40.54 vs. \qty{44.20}{\mu_B/u.c.}) despite its high anisotropy, diverges in several ways: MCA energy decreases under both compressive and tensile strains. Furthermore, the peak in the Fe(4p) density remains above the Fermi level regardless of strain, while the Fermi surface topology is significantly different, with more features in the $xy$-plane. This suggests that the specific peak alignment discussed above is more strongly tied to high magnetic moment than high anisotropy in \fn. Regardless, the existence of these peaks ties back to the partial localization of Fe electronic states observed in \fn\cite{wang2012} and echoes observations in other Fe-based materials where tuning occupation of localized vs. itinerant Fe states strongly affects magnetic order.\cite{li2026}  These results underscore the complex, correlated character of \fn\ and its highly anisotropic magnetic interactions.


\section{Conclusion}

In summary, the magnetic properties of \afn\ were systematically investigated within the GGA+$U$ framework. The comprehensive survey across the Hubbard parameter space revealed complex, non-linear dependencies in $M_s$ and especially MCA energy. Investigation of specific cases of interest showed increased $U$ on Fe sites generally leading to charge transfer from Fe to N while increasing local magnetic moments. The projected DOS contained notable peaks of Fe(4p) occupation close to the Fermi energy, consistent with a van Hove singularity. These peaks were more strongly affected by in-plane strain than by precise Hubbard $U$ parameters, implying an intrinsic effect connected to local crystal field splitting.

Our analysis thus suggests that ``giant'' magnetization in \fn\ is intrinsically coupled to this critical electronic instability, with the high-moment phase in a precarious energy landscape where the balance between local and itinerant Fe(3d/4p) states is easily perturbed, contributing to the material’s structural and phase metastability. Further investigation of these features and potential resultant emergent many-body instabilities in high-$M_s$ \fn\ at the GGA+$U$ level and beyond is thus well warranted. Furthermore, optimizing \fn\ for practical applications may require moving beyond simple moment or anisotropy maximization to directly address Fermi surface topology and mitigate the underlying instabilities. These findings underscore the complex, correlated nature of iron nitrides and define the electronic constraints within which stable, high-performance permanent magnets must be engineered.

\begin{acknowledgments}
Authors deeply appreciate the Minnesota Supercomputing Institute (MSI) for providing computational resources used for the DFT calculations (see http://www.msi.umn.edu), as well as the Robert F. Hartmann Endowed Chair.
Peter Stoeckl acknowledges partial financial support from Niron Magnetics. 

Fig. 1 was produced using Mathematica, Version 11.3.0, by Wolfram Research, Inc.\cite{Mathematica2018}

\bigskip
Dr. Jian-Ping Wang has equity and royalty interests in and serves on the Board of Directors for Niron Magnetics Inc, a company involved in the commercialization of Fe16N2 magnet. The University of Minnesota also has equity and royalty interests in Niron Magnetics Inc. These interests have been reviewed and managed by the University of Minnesota in accordance with its Conflict of Interest policies.
\end{acknowledgments}

\bibliography{references}


\appendix
\clearpage

\onecolumngrid
\renewcommand{\thefigure}{S\arabic{figure}}  
\renewcommand{\thetable}{S-\Roman{table}}  
\setcounter{figure}{0}
\setcounter{table}{0}

\section*{Supplemental Material}

\subsection*{Previous observations of MCA energy in \fn}

\begin{table}[ht]
 \caption{\label{t:sm_prev_mca} Summary of selected previous observations of MCA energy in \fn, both experimental and theoretical. Lattice constants chosen in theoretical results are included for context (values in parentheses give the usual experimental values, as given in Ref. \cite{jack1951}).
}
 \begin{ruledtabular}
  \begin{tabular}{ccc cc r c}
  \multicolumn{2}{c}{\multirow{2}{*}{Paper}} & \multirow{2}{*}{Ref.} & \multicolumn{2}{c}{Lattice Const. (\AA)} & \multicolumn{1}{c}{$K_u$} & \multirow{2}{*}{Notes} \\
  \cline{4-5}
  &&& $a$ & $c$ & \multicolumn{1}{c}{($10^5$ erg/cc)} & \\
  \hline
  \multirow{6}{*}{(exp.)} & Takahashi 1999 & \cite{takahashi1999} & (5.72) & (6.29) & 200 & $K_u$ taken as $K_1+K_2$ \\
  & Kita 2007 & \cite{kita2007} & (5.72) & (6.29) & 44 & nanoparticles \\
  & Ji 2011b & \cite{ji2011b} & (5.72) & (6.29) & 100 & \\
  & Ogawa 2013 & \cite{ogawa2013} & (5.72) & (6.29) & 96 & nanoparticles \\
  & Li 2019 & \cite{li2019} & (5.72) & (6.29) & 190 & \\
  & Liu 2020 & \cite{liu2020} & (5.72) & (6.29) & 70 & \\
  \hline
  \multirow{10}{*}{(theo.)} & Ke 2013 & \cite{ke2013} & 5.72 & 6.29 & 144 / 131 & LDA / GGA (resp.) \\
  & Zhou 2014 & \cite{zhou2014} & 5.68 & 6.23 & 105 / 146 & GGA / GGA+$U$ \\
  & Zhao 2016 & \cite{zhao2016} & 5.68 & 6.22 & 64 & GGA (pure \fn) \\
  & Han 2019 & \cite{han2019} & 5.69 & 6.22 & 86 & GGA (pure \fn) \\
  & Li 2019 & \cite{li2019} & 5.71 & 6.28 & 50 / 160 & LDA / LDA+$U$ ($U = \qty{4}{eV}$) \\
  & Islam \& Borah 2020 & \cite{islam2020} & 5.696 & 6.185 & 81 & GGA+$U$ (pure \fn) \\
  & Sun 2020 & \cite{sun2020} & 5.68 & 6.22 & 68 & GGA (pure \fn) \\
  & Ochirkhuyag 2021a & \cite{ochirkhuyag2021a} & 5.69 & 6.25 & 73 & GGA (pure \fn) \\
  & Kota \& Sakuma 2023 & \cite{kota2023} & 5.68 & 6.23 & 61 & GGA \\
  & Sakuma 2023 & \cite{sakuma2023} & 5.72 & 6.29 & 100 & TB--LMTO (pure \fn) \\
  \end{tabular}
 \end{ruledtabular}
\end{table}

\subsection*{Summary of previous results with $U_{4d} = \qty{0.9}{eV}$ fixed}

The majority of the results in this section were previously published in Ref. \cite{stoeckl2023}: Following the model proposed by Ji \textit{et al.} in 2010,\cite{ji2010} Hubbard $U$ was fixed at a low value of \qty{1.0}{eV} on the Fe$4d$ site while simultaneously varying $U$ on the Fe$4e$ and Fe$8h$ sites from \qty{0.5}{eV} up to \qty{4.0}{eV}. In calculations with spin-orbit coupling (SOC), the Hund's coupling parameter $J$ was taken to be $J=U/10$ on all three Fe sites. For the preceding structural relaxation calculations, the effective parameter $U_{eff}$ was set equal to $U-J$ from the SOC calculations, resulting in $U_{eff}$ ranging from \qty{0.45}{eV} to 
\qty{3.60}{eV} on the Fe$4e$/Fe$8h$ sites.

To minimize the potential discrepancies between the structural relaxation calculations (using $U_{eff}$) and the magnetic property calculations (using $U$ and $J$), a second systematic survey was conducted with $J=0$ throughout, so that the $U$ value used in the SOC calculations matched $U_{eff}$ used for the structural relaxation (as $U_{eff} = U - J$ now simplifies to $U_{eff} = U$). By choosing the $U$ values in this second SOC series to equal the $U_{eff}$ values from the first, the same set of relaxed structures could be used for both. (Test calculations confirmed that non-SOC ground state energies calculated with $U_{eff}$ matched perfectly with those calculated using $U_{eff} = U$ when $J=0$, validating this approach.)

\begin{figure}
 \includegraphics{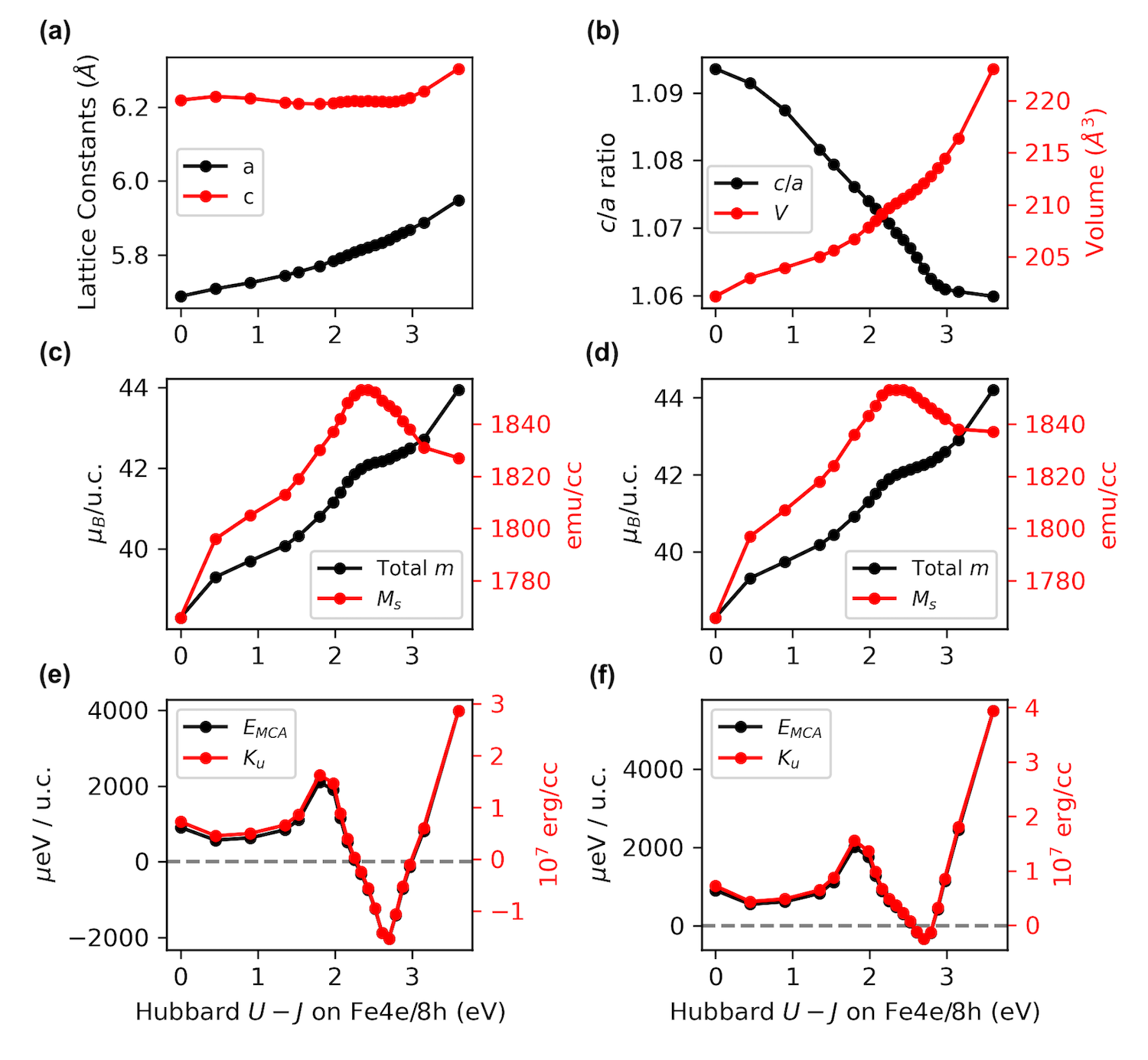}%
 \caption{\label{f:sm_uj_data} Lattice and magnetic parameters as function of $U-J$ on Fe$4e$/$8h$. (a) Lattice parameters $a,c$. (b) $c/a$ ratio, and volume of the bct unit cell. (c--d) Total magnetic moment per unit cell and equivalent $M_s$: (c) for $J = U/10$, (d) for $J = 0$. (e--f) MCA energy and equivalent $K_u$: (e) for $J = U/10$, (f) for $J = 0$. $U-J = 0$ corresponds to the GGA calculation throughout.
}
\end{figure} 

Detail on relaxed crystal structures is given in \Cref{f:sm_uj_data}(a-b). Increasing $U-J$ on Fe$4e$/$8h$ mainly leads to expansion in the $ab$ plane, with $c$ increasing significantly only for $U-J > \qty{3}{eV}$ (at which point $a$ well exceeds its experimental value) --- thus increasing unit cell volume while reducing $c/a$.

Magnetic properties for the first series ($J=U/10$) are shown in \Cref{f:sm_uj_data}(c,e). While magnetic moment increases with increased $U-J$, saturation magnetization $M_s$ peaks around $U-J \sim \qty{2}{eV}$ due to the lattice expansion. However, even the peak value of \qty{1850}{emu/cc} remains well below several recent experimental observations of  $M_s > \qty{2000}{emu/cc}$.\cite{wang2012,li2019,hang2020} Magnetocrystalline anisotropy, however, behaves less linearly: first decreasing slightly compared to pure GGA, then increasing with $U-J$ up to $\sim\qty{2}{eV}$. As $U-J$ increases further, MCA energy decreases again, including a change in sign (from out-of-plane to in-plane easy axis), before increasing back into positive values well above experimental observations. Closest agreement to experiment is either around $U-J \sim \qty{2}{eV}$ or $U-J \sim \qty{3}{eV}$.

In the second series, with $J=0$ throughout, the trends are similar (cf. \Cref{f:sm_uj_data}(d,f)), especially for magnetization. For anisotropy, however, the region of decreased or negative MCA energy ($U-J \sim \qtyrange{2}{3}{eV}$) is greatly reduced, only slightly dipping below 0, and increasing even further positive at higher $U-J$. This suggests that this nonlinear behavior cannot be entirely explained by the “effective strain” resulting from using $U_{eff}$ to relax crystal structures rather than the explicit $U,J$ used in magnetic property calculations.

\begin{figure}
 \includegraphics[scale=0.75]{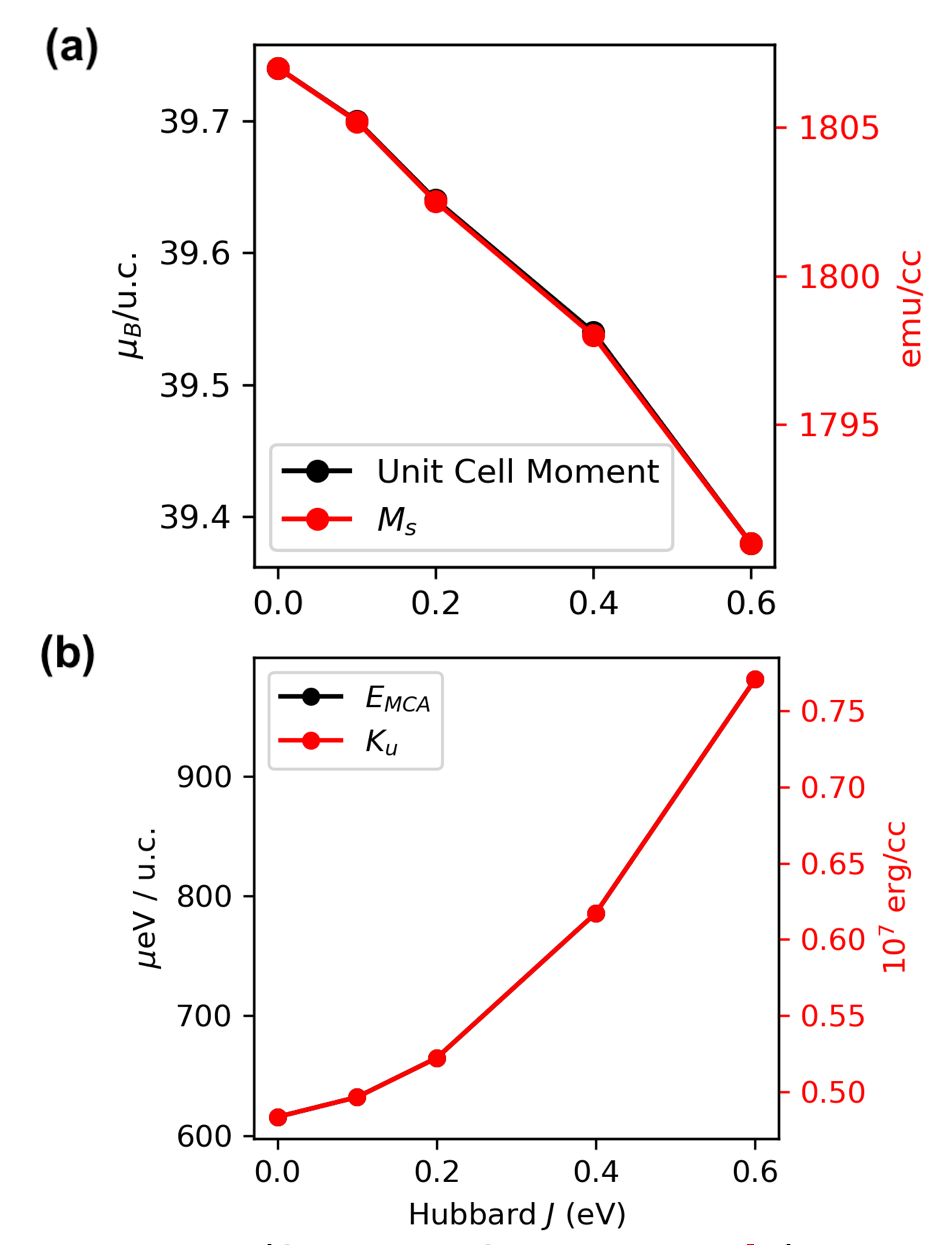}%
 \caption{\label{f:sm_j_data} Magnetic parameters as function of $J$, for $U-J = \qty{0.9}{eV}$ and $J$ identical across all Fe sites. (a) Total magnetic moment per unit cell and equivalent $M_s$. (b) MCA energy and equivalent $K_u$.
}
\end{figure}

Finally, to isolate the effect of the exchange parameter $J$, we fixed $U-J = \qty{0.9}{eV}$ on all Fe, and varied the value of $J$ from \qty{0}{eV} up to \qty{0.6}{eV}. (As $U-J$ was fixed, the same relaxed structure was used in all cases.) The results were comparatively straightforward: increasing $J$ with $U-J$ fixed leads to significant, monotonic increase in MCA, from \qty{616}{\mu{}eV/u.c.} up to \qty{981}{\mu{}eV/u.c.}, with only a slight reduction in total magnetic moment (\qty{39.74}{\mu_B/u.c.} to \qty{39.38}{\mu_B/u.c.}) — compare \Cref{f:sm_j_data}.

\subsubsection*{Linear-response calculations}

For comparison, a linear-response calculation (following the methodology of Cococcioni and de Gironcoli\cite{cococcioni2005}) yielded the following $U_{eff}$ parameters: $U_{4e} = \qty{4.26}{eV}$, $U_{8h} = \qty{4.18}{eV}$, $U_{4d} = \qty{3.85}{eV}$; note that here $U_{4d}$ is still smallest, though greater than $U_{4d} = \qty{0.9}{eV}$ used elsewhere in these data. Unfortunately, upon relaxing the crystal structure with these $U_{eff}$, the resulting lattice parameters ($a = \qty{6.094}{\textup{\AA}}$, $c = \qty{6.503}{\textup{\AA}}$) were considered too strongly distorted to yield a reliable estimate of $\Delta E_{MCA}$.

\subsection*{Extension of data set for $U_{4e}\neq U_{8h}$ at higher values of $U_{4d}$}

\begin{figure}[h]
 \includegraphics{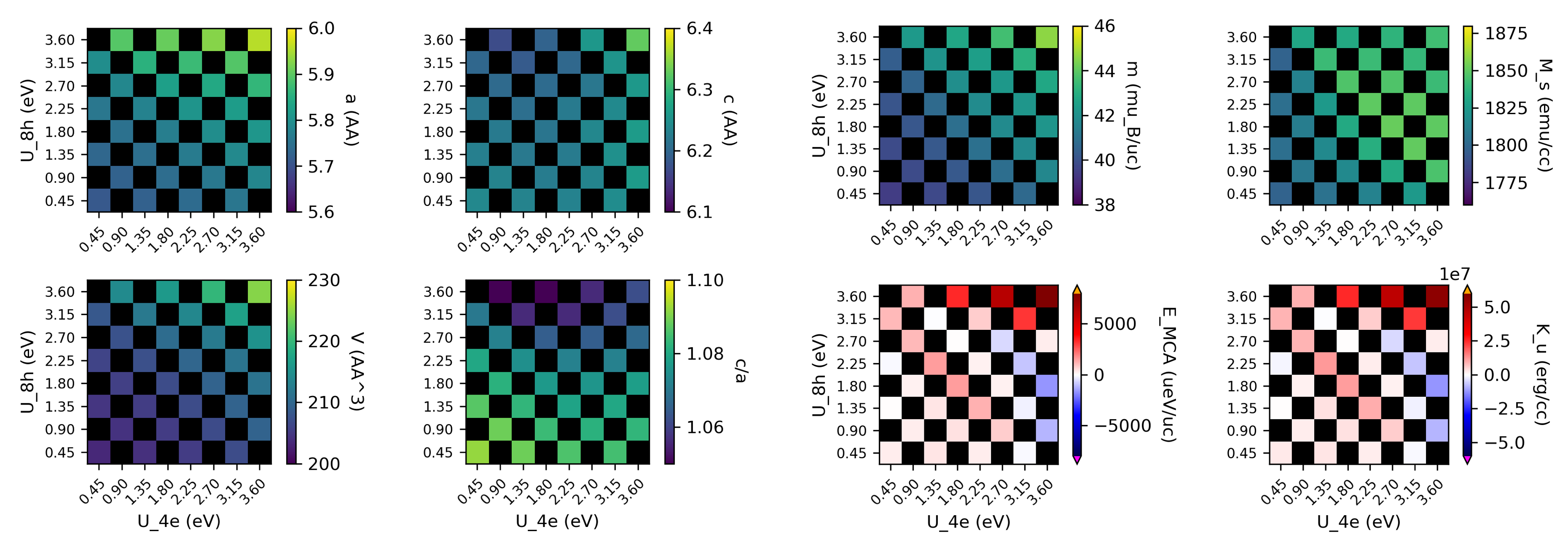}%
 \caption{\label{f:sm_set_uneq2} Lattice and magnetic parameters of \fn under GGA+$U$, with $U_{4e}\neq U_{8h}$ varying independently and $U_{4d} = \qty{1.80}{eV}$ fixed. Left: lattice parameters; right: magnetic parameters. Black squares indicate missing data (calculations skipped).
}
\end{figure}

\begin{figure}[h]
 \includegraphics{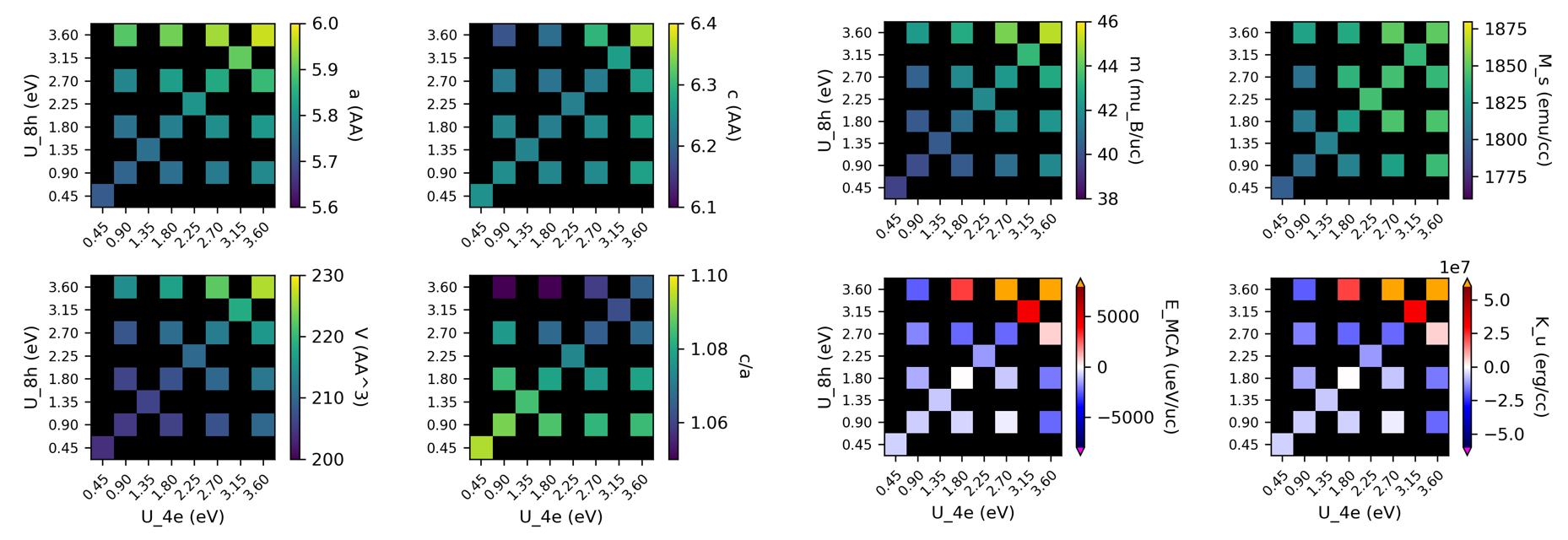}%
 \caption{\label{f:sm_set_uneq3} Lattice and magnetic parameters of \fn under GGA+$U$, with $U_{4e}\neq U_{8h}$ varying independently and $U_{4d} = \qty{2.70}{eV}$ fixed. Left: lattice parameters; right: magnetic parameters. Black squares indicate missing data (calculations skipped).
}
\end{figure}

\subsection*{Crystal-field-split pDOS}

In the crystal structure of \afn (I4/mmm), the local symmetries of Fe sites are reduced from the overall $D_{4h}$. Specifically, the Fe$4e$ site has $C_{4v}$ (4mm) point group symmetry; the Fe$4d$ site, $C_{2v}$ (mm2); and Fe$8h$ is reduced to $C_{1h}$ (m).  The p orbitals of Fe (derived from atomic $\mathrm{p_x, p_y, p_z}$) split according to these local symmetries due to the crystal field. In the $C_{4v}$ symmetry, those orbitals split into a doubly degenerate $E$ set ($\mathrm{p_x, p_y}$) and a non-degenerate $A_1$ ($\mathrm{p_z}$). The $\mathrm{p_z}$ orbital transforms as the $A_1$ irrep along the tetragonal $c$-axis, while $\mathrm{p_x}$ and $\mathrm{p_y}$ remain degenerate due to the fourfold rotation ($C_4$) and mirror planes. Typically, $\mathrm{p_z}$ ($A_1$) is lower in energy than the $E$ pair in a tetragonal field, but this depends on the axial ligand field strength (\textit{e.g.}, compression along $c$ enhances $A_1$ stabilization). Since these sites are closer to N atoms, $\mathrm{p_z}$ hybridization with N $\mathrm{pz}$ may be prominent, affecting covalent bonding between atoms (see main text, \Cref{f:structure}). Similarly, under $C_{1h}$, $\mathrm{p_x}$ and $\mathrm{p_y}$ ($A'$) split from $\mathrm{p_z}$ ($A''$). In the $C_{2v}$ symmetry, by contrast, all three p orbitals are fully split into non-degenerate singlets: $A_1$ ($\mathrm{p_z}$), $B_1$ ($\mathrm{p_x}$), and $B_2$ ($\mathrm{p_y}$). No degeneracy remains due to the lower symmetry (only $C_2$ rotation and two mirror planes). The orbital energy ordering is highly anisotropic and $\mathrm{p_x}$ ($B_1$) and $\mathrm{p_y}$ ($B_2$) often differ further based on equatorial distortions, with $\mathrm{p_z}$ ($A_1$) typically lowest. These equatorial Fe atoms experience more lateral strain, leading to greater p-orbital involvement in Fe-Fe bonding.

In turn, the nitrogen atoms occupy the $2a$ Wyckoff position, which has a local point group symmetry of $D_{4h}$ (4/mmm), characterized by a fourfold rotation axis along the $c$-direction, horizontal and vertical mirror planes, and inversion. In the ideal octahedral ($O_h$) environment, the three 2p orbitals are triply degenerate ($T_{1u}$ irrep). The $D_{4h}$ distortion splits them into $A_{2u}$ ($\mathrm{p_z}$)---non-degenerate and oriented along the tetragonal $c$-axis---and a doubly degenerate $E_u$ pair ($\mathrm{p_x, p_y}$), oriented in the basal $xy$-plane. The $\mathrm{p_z}$ orbital has $\sigma$-symmetry, strongly interacting with the closer apical Fe$4e$ atoms, while the $\mathrm{p_x}$ and $\mathrm{p_y}$ orbitals have $\pi$-symmetry, interacting more equally with all six Fe atoms but weaker overall due to longer equatorial bonds.

\begin{figure}
 \includegraphics{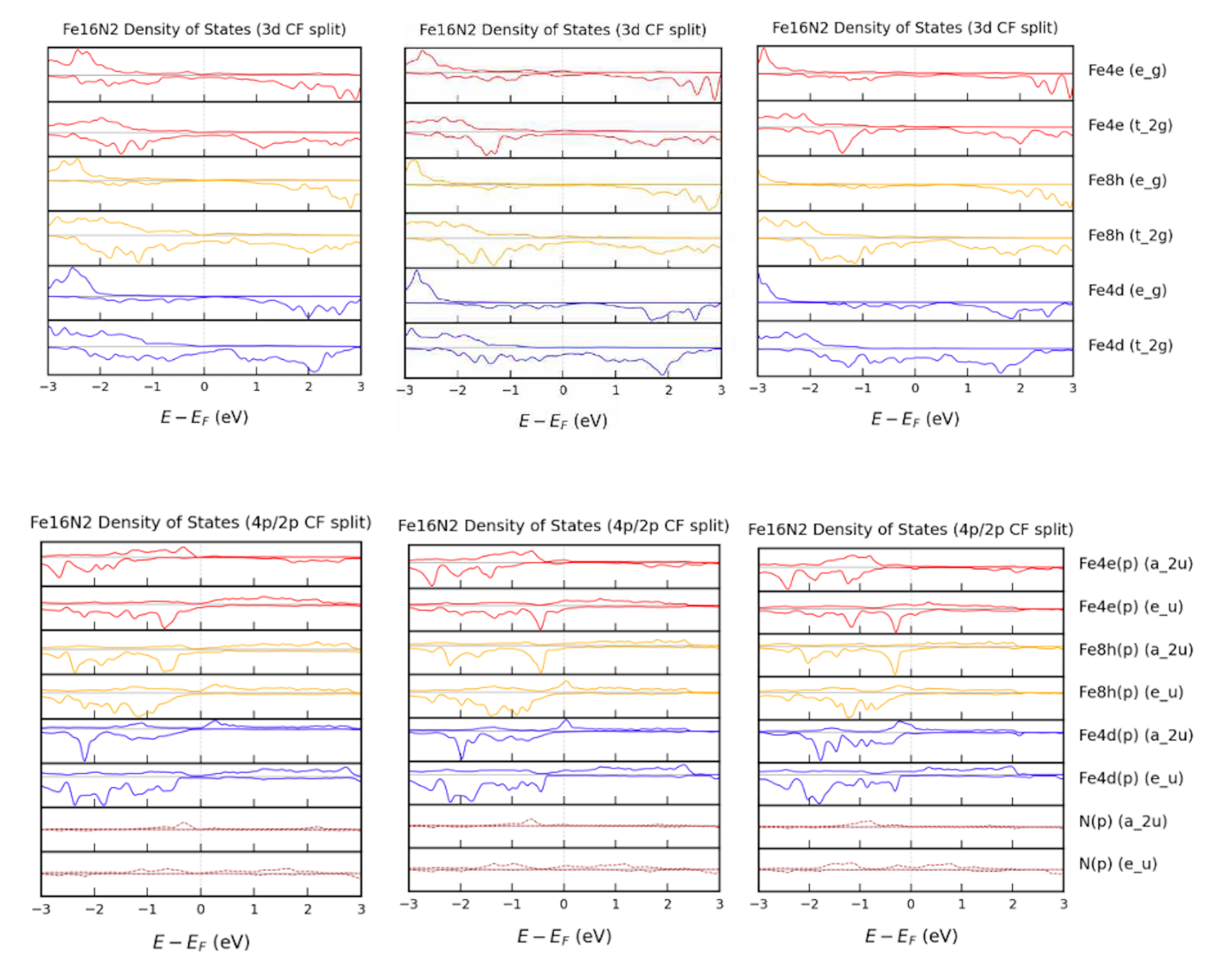}%
 \caption{\label{f:sm_crystal_field} Case \textbf{A4} ($U_{4e} = U_{8h} = \qty{3.15}{eV}$, $U_{4d} = \qty{2.25}{eV}$): crystal-field split pDOS near $E_F$.\\
Top: Fe(3d), site-projected. Bottom: Fe(4p)/N(2p) [N on different y-axis scale].\\
Left: $-4\%$ $ab$-strain; Center: 0 strain; Right: $+4\%$ $ab$-strain. Symmetry labels correspond to $D_{4h}$ splitting: $e_g$ ($z^2/x^2-y^2$), $t_{2g}$ ($xz/yz/xy$) for 3d; $a_{2u}$ ($z$), $e_u$ ($x/y$) for 4p.
}
\end{figure}

The above symmetries are preserved under biaxial in-plane strain (varying $a=b$ together). If the in-plane axes are not perfectly equivalent, applying strain to the lattice cell distorts the structure and introduces slight orthorhombicity, lowering $C_{4v}$ to $C_{2v}$ (or even $C_{1h}$ if severe), thus $\mathrm{p_x/p_y}$ ($E$) splits into $B_1$ ($\mathrm{p_x}$) and $B_2$ ($\mathrm{p_y}$). This phenomenon is typical in \fn\ thin films, often stemming from the $\sim 4\%$ lattice mismatch with \ce{MgO} substrate. This effect can be also easily modulated by mechanical strain, crystal distortion, or doping with other elements.

\clearpage

\subsection*{Additional figures for case A4}

\begin{figure}[h]
 \includegraphics{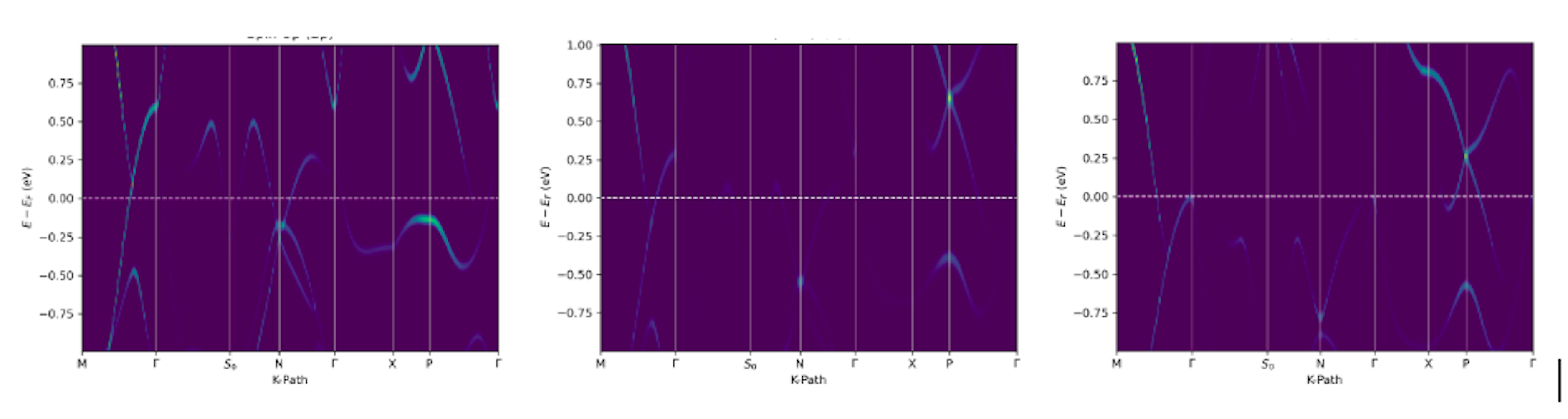}%
 \caption{\label{f:sm_pdos_n2p} \emph{Spin-up} $k$-resolved pDOS projected on N(2p) charge density (color).\\
 Left: $-4\%$ $ab$-strain; Center: 0 strain; Right: $+4\%$ $ab$-strain.
}
\end{figure}

\begin{figure}[h]
 \includegraphics{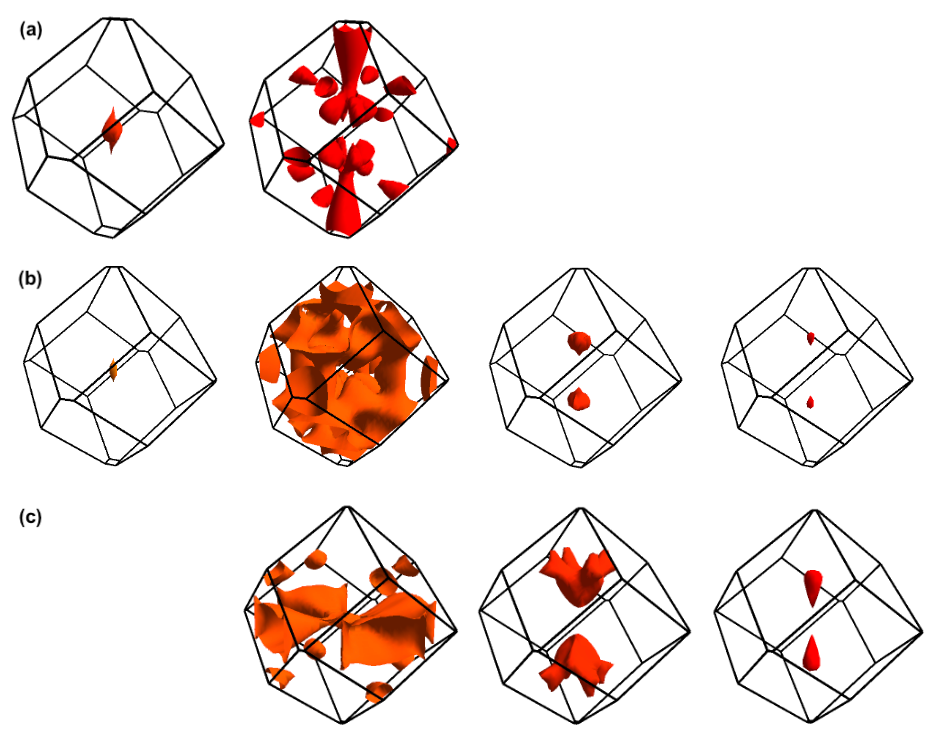}%
 \caption{\label{f:sm_fs_strain} Fermi surfaces for case \textbf{A4} --- \emph{spin up}, split by band (left to right); (a) $-4\%$ strain; (b) 0 strain; (c) $+4\%$ strain.
}
\end{figure}

\clearpage

\subsection*{Repetition of Figs. 5--6 for cases A5 / B3}

\begin{figure}[h]
 \includegraphics{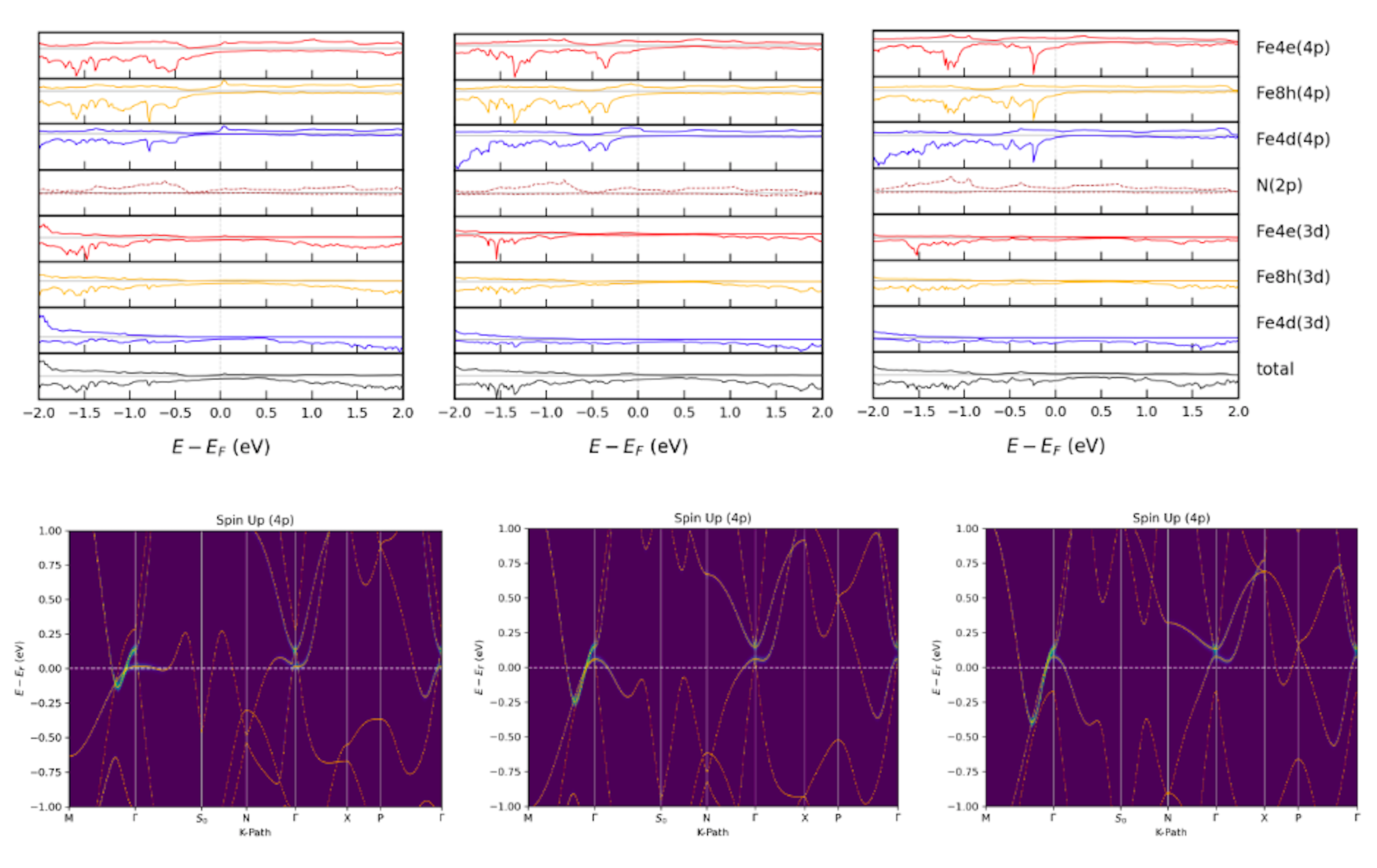}%
 \caption{\label{f:sm_case_a5} Case \textbf{A5} ($U_{4e} = U_{8h} = \qty{3.60}{eV}$, $U_{4d} = \qty{0.90}{eV}$).\\
Top: Orbital-resolved pDOS near the Fermi energy. Bottom: \emph{Spin-up} band structure (orange) near the Fermi energy, superimposed on $k$-resolved pDOS projected on Fe(4p) charge density (color). \\
Left: $-4\%$ $ab$-strain ($M=\qty{42.44}{\mu_B/u.c.}$, $\Delta E_{MCA}=\qty{+1248}{\mu{}eV/u.c.}$); \\
Center: 0 strain ($M=\qty{44.20}{\mu_B/u.c.}$, $\Delta E_{MCA}=\qty{+5475}{\mu{}eV/u.c.}$); \\
Right: $+4\%$ $ab$-strain ($M=\qty{45.66}{\mu_B/u.c.}$, $\Delta E_{MCA}=\qty{+5541}{\mu{}eV/u.c.}$)
}
\end{figure}

\begin{figure}[ht]
 \includegraphics{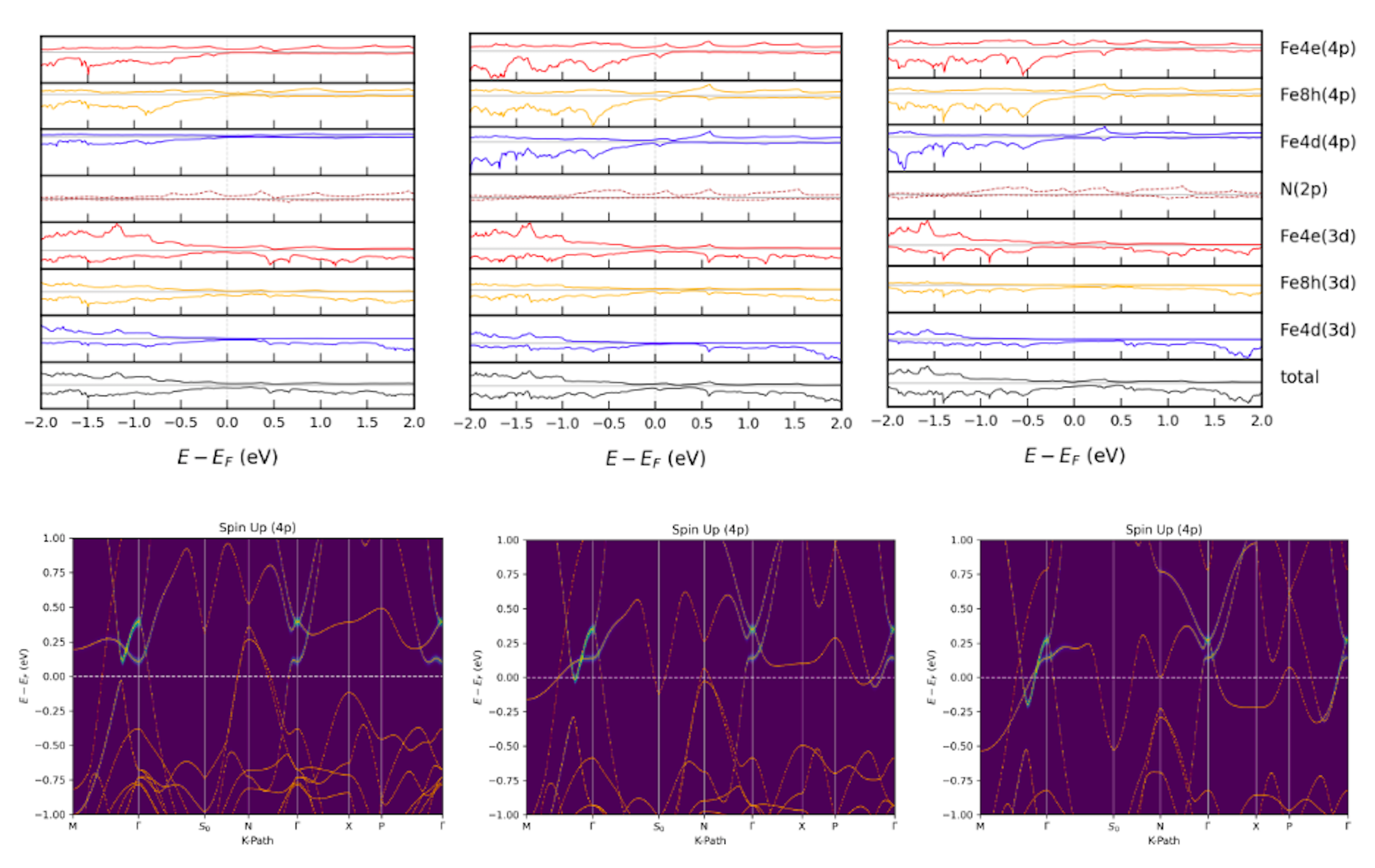}%
 \caption{\label{f:sm_case_b3} Case \textbf{B3} ($U_{4e} = \qty{0.45}{eV}$, $U_{8h} = \qty{3.15}{eV}$, $U_{4d} = \qty{0.90}{eV}$).\\
Top: Orbital-resolved pDOS near the Fermi energy. Bottom: \emph{Spin-up} band structure (orange) near the Fermi energy, superimposed on $k$-resolved pDOS projected on Fe(4p) charge density (color). \\
Left: $-4\%$ $ab$-strain ($M=\qty{39.26}{\mu_B/u.c.}$, $\Delta E_{MCA}=\qty{+1772}{\mu{}eV/u.c.}$); \\
Center: 0 strain ($M=\qty{40.54}{\mu_B/u.c.}$, $\Delta E_{MCA}=\qty{+4733}{\mu{}eV/u.c.}$); \\
Right: $+4\%$ $ab$-strain ($M=\qty{42.22}{\mu_B/u.c.}$, $\Delta E_{MCA}=\qty{+1052}{\mu{}eV/u.c.}$)
}
\end{figure}

\begin{figure}
 \includegraphics{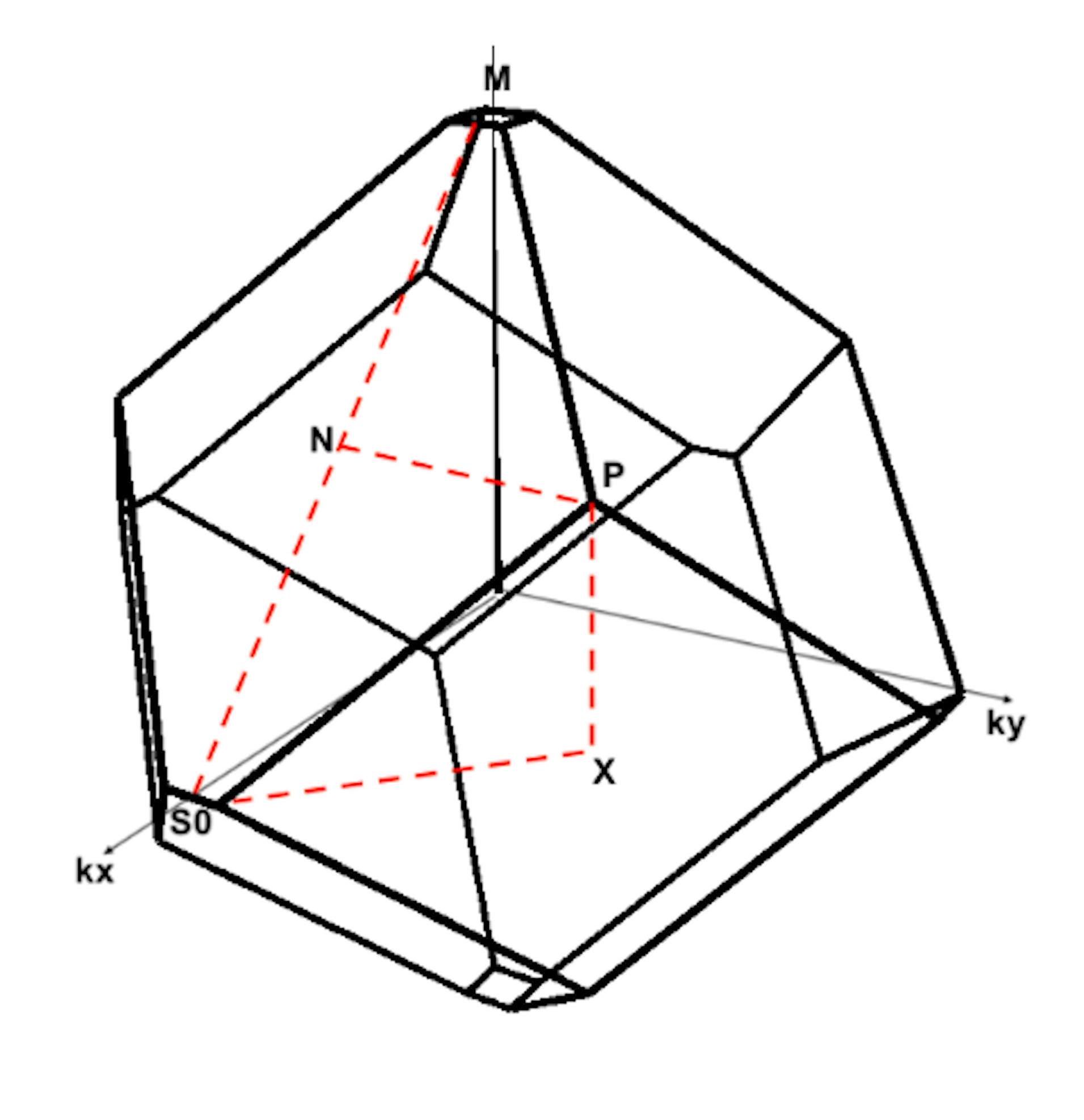}%
 \caption{\label{f:sm_bz} Sketch of significant $k$-points in the Brillouin Zone for bct \fn
}
\end{figure}

\end{document}